# Mapping giant magnetic fields around dense solid plasmas by high resolution magneto-optical microscopy

Jaivardhan Sinha, Shyam Mohan, and S. S. Banerjee[a]

Department of Physics, Indian Institute of Technology, Kanpur-208016, U. P., India

Subhendu Kahaly and G. Ravindra Kumar[b]

Tata Institute of Fundamental Research, 1 Homi Bhabha Road, Mumbai 400 005, India

We investigate distribution of magnetic fields around dense solid plasmas generated by intense p-polarized laser (~$10^{16}$ Wcm$^{-2}$, 100 fs) irradiation of magnetic tapes, using high sensitivity magneto optical microscopy. We present evidence for giant axial magnetic fields and map out for the first time the spatial distribution of these fields. By using the axial magnetic field distribution as a diagnostic tool we uncover evidence for angular momentum associated with the plasma. We believe this study holds significance for investigating the process under which a magnetic material magnetizes or demagnetizes under the influence of ultrashort intense laser pulses.

[a] email: satyajit@iitk.ac.in

[b] email: grk@tifr.res.in

Contemporary fascination with ultralarge, megagauss (*MG*) scale magnetic fields is inspired by (a) their unique influence on fundamental properties of condensed matter systems at low temperatures on one end and (b) their role in hot, dense, highly ionized complex plasmas that mimic astrophysical scenarios at the other. Fields as large as 700 MG - the largest on earth- found in intense picosecond laser produced solid plasmas[1] offer tantalizing possibilities to simulate the behavior of neutron stars in the laboratory[2]. At low temperatures, giant magnetic fields might enable the exploration of exotic quantum fluctuation driven phase transitions[3,4] and new phenomena like the now well known fractional quantum Hall effect[4] associated with 2D electron gas in semiconductors. Pulsed, giant magnetic fields also impact current efforts to realize 'fast ignition' of laser fusion[5] as they crucially modify relativistic electron transport, a key physics issue of this scheme. We have recently demonstrated *MG* field pulses of subpicosecond duration and extracted the hitherto intractable electrical conductivity from them[6]. Knowledge of the spatial and temporal behavior of large magnetic fields is crucial for applying them in fundamental study and manipulation of electronic properties of materials.

Strong azimuthal magnetic fields (~ O(100) *MG*) are generated in dense plasmas using the explosive ionization of a solid target by intense, ultrashort laser pulses[1] which generate megaamperes of relativistic 'hot' electron currents directed into the target.

Magneto optical (MO) techniques (Faraday and Cotton Mouton) have been used for probing the *azimuthal* magnetic field created by resonance absorption (RA) in plasmas[6]. Briefly, RA involves the generation of electron plasma waves by the absorption of a p-polarized laser, incident at an oblique angle to the target normal[7]. When these waves are damped, they generate a hot electron current, directed along the target normal, which in turn gives rise to azimuthal fields[6,7]. An issue of fundamental significance is whether the massive azimuthal fields couple in some way to the plasma thereby producing another component of the magnet field. To date, axial magnetic fields in intense laser produced plasmas have been demonstrated only in experiments using normally incident circularly polarized laser light[8]. In this paper we not only present clear evidence for large *axial* magnetic fields generated by *p-polarized* light excitation (i.e. under RA conditions) but also obtain unique information about the distribution of these axial fields around the irradiated spot. We employ the high sensitivity magneto optical microscopy (MOM) technique, which has been used extensively in recent studies of condensed matter systems[9,10]. The Faraday rotation based MOM which is *blind* to the azimuthal magnetic field enables imaging of only the axial magnetic field distribution. The axial field distribution is imprinted on a magnetic recordable media (magnetic tape) after it is irradiated with 100 femtosecond p-polarized laser pulses at intensities in the range of $10^{15}$-$10^{16}$ W cm$^{-2}$. Dramatic changes in the magnetic domain patterns are observed on the tape after laser irradiation. By analyzing the distribution of axial magnetic fields we infer the presence of azimuthal circulating currents in the plasma thereby indicating broken rotational symmetry.

We irradiated the magnetic tape mounted in a vacuum chamber with p-polarized pulses from a table top terawatt laser (806 nm, 10Hz and 100 fs)[6] at $45^0$ angle of incidence. Each laser shot irradiated a fresh spot on the tape. The tape consists of nanometer sized $\gamma$ - $Fe_2O_3$ particles suspended in a polymeric, non-magnetic matrix[11,12]. While the magnetic properties of the nanometer sized ferromagnetic particles are of current interest[13], we will be concerned only with the bulk magnetic response of the ferromagnetic tapes, as the MOM imaging technique measures the magnetization response averaged over a large number of nanoparticles. Typical size of the tapes used was *7 x 4* mm$^2$ and thickness O(*100 nm*)[11,12]. Energy dispersive X-ray spectroscopic (EDAX) analysis in a scanning electron microscope (Model QUANTA-200, from FEI) indicated a layer of $Fe_2O_3$ constituting the tape and the magnetic properties of the tape were investigated using a commercial vibrating sample magnetometer (VSM) (Model EV7-VSM; ADE Technologies), with a sensitivity of *$10^{-5}$* e.m.u. After laser irradiation, the tapes were transferred from the irradiation chamber to the MOM setup. Faraday rotation[14] is the basis of the MOM set up which has a reflecting type polarizing microscope (Carl Zeiss, model: Axio Tech Vario) and a Peltier cooled Charge Coupled Device (CCD) camera (Andor, iXon, 512 x 512 pixels, 16 bit, 85% quantum efficiency at 550 nm). The ray diagram in our MOM setup is shown in Fig.1(a). During the experiment, a material with high

Verdet's constant (*v*) (ferrite garnet (FGF) film, marked V in Fig.1(a)) is placed in direct contact with the irradiated tape (T, in Fig.1(a)). The linearly polarized incident light (blue ray in Fig.1(a)) undergoes Faraday rotation upon reflection from V. The angle of Faraday rotation $\theta = B_z v(2d)$, where d is the thickness of the FGF film and $B_z$ is the local field perpendicular to the tape surface (axial) placed underneath the FGF film. The high sensitivity CCD camera captures the Faraday rotated light intensity variations which correspond to the spatial distribution of the axial magnetic field viz., $B_z(x, y)$ and the local intensity of the image (i.e., the gray level) is directly related to the value of $B_z(x, y)$ and its direction. In this mode of operation the magnetic field sensitivity is 0.5 G (though much higher sensitivity is possible[10]) with a spatial resolution of 0.5 µm.

As is well known, information in a magnetic recordable tape is encoded via a unique arrangement of magnetic domains on the tape. Fig.1(b), shows the MOM image of magnetic domains in a region of the tape where data are encoded. In the MOM image, a single magnetic domain is characterized by bright and dark lines with gray in between (see area inside the dotted lines in the inset of Fig.1(b)). The white and black lines are magnetic domain edges corresponding to $+B_z$ pointing out of and $-B_z$ pointing into the tape surface, respectively. The setup is calibrated by measuring $\theta$ for known magnetic field values. Using this calibration scheme the $B_z(x, y)$ across the tape surface can be inferred. Fig.1 shows the $B_z(x)$ profile measured across the solid horizontal black line on the tape (Fig.1(b)).

To determine the pristine nature of the magnetic fields generated in our experiment, a region of the tape devoid of predefined magnetic domains was chosen for irradiation viz., regions *devoid of* black-gray-white patterns shown in Fig.1(b). The tape was irradiated at six locations (arranged on the vertices of a hexagon) with single shots from the femtosecond laser pulse with energy *6.0 - 6.4* mJ. The spot size of the irradiating laser pulse was *20* $\mu$m. The heating generated by the laser irradiation is highly local as the medium of the tape is a non conducting polymer which aids in the creation of the plasma at the irradiation site. Fig. 2(a) shows the MOM image of the six irradiated spots. Fig. 2(b) shows the line scan ($B_z(x)$) across the dark line in Fig.2(a). It is seen that $B_z(x)$ changes sign across the irradiated spot, viz., the field varies from *+70* G on the periphery of the spot to *–70* G in the center of the spot. We will discuss the significance of the sign change of $B_z$ subsequently. The smooth and continuous behavior of the axial magnetic field distribution across the entire irradiated region implies that the laser irradiation does not destructively damage the magnetic layer of the tape. In Fig.2(b) we find that $B_z(x)$ reaches a maximum of *+70* G on one edge of the irradiated spot and in a diametrically opposite region on the periphery it is only *+20* G, indicating a non-uniform local field profile ($B_z(x, y)$) around the spot. In Fig.2(c) we show a high magnification MOM image of a single irradiated spot (region within the dashed, black circle in Fig.2(a)). Fig. 2(d) is the contour plot of the magnetic field distribution ($B_z(x, y)$) deduced from the MO image in Fig.2(c), where each color represents uniform $B_z(x, y)$ values. From Fig.2(d) it is clear that the axial magnetic field ($B_z$) generated by the laser pulse irradiation has a complex and highly nonuniform spatial distribution around the irradiated spot. The above features are common to all the six irradiated spots.

Next, we show in Fig.3(a), an MOM image in a prerecorded region ( i.e. information coded and domain aligned as in Fig.1) of the tape irradiated with high intensity laser pulses. The region around the laser irradiation is encircled and the $B_z(x, y)$ distribution within it is similar to that in Fig.2(c), where {$B_z < 100$ G; for all (x, y)}. It is interesting to note that sharp lines observed with MOM on the prerecorded tape before irradiation cf. Fig.1(b), have completely faded away after the irradiation, cf. Fig.3(a). Can this effect be produced with magnetic fields of O(*100)* G? Shown in Fig. 3(b) is the ferromagnetic hysteresis loop of the magnetic tape recorded on a VSM. The magnetization response of the tape *saturates* beyond 0.*3* Tesla. In separate measurements with an electromagnet, we observed that fields as large as 0.3T had to be applied before the unirradiated tapes showed the effect of fading of domains. We therefore infer that such fading of domains as in Fig.3(a) is due to the large magnetic fields overwhelming the domain patterns on the tape and orienting all the domains in the same directions, thereby diminishing the contrast (bright and dark lines) between different domains. The lines are affected at distances as large as *200* μm from the irradiated spot, from which we conclude that at distances as large as 20 times the radius of the irradiating laser beam (*10* μm) magnetic fields of atleast 0.3 T are created (note that saturation of magnetization limits this estimate). We believe that inevitable heating around the laser irradiated spot causes decay in the magnetization from the much larger saturation magnetization down to ~ 100 G which is detected on the tape by MOM.

The decay of magnetization in the tape is due to heating effects which can be estimated by VSM measurements of the temperature ($T$) dependence of the remnant magnetization ($M_{rem}$) of the tape. To measure $M_{rem}$, a magnetic field of 1 T which saturates the magnetization of the tape is applied and subsequently switched off. A temperature dependence of the type $M_{rem}(T) = M_{rem}(T=0) + T.(dM_{rem}/dT)$, with $M_{rem}(T=0) = 1.5$ T and $dM_{rem}/dT = -26.4$ G/°K, is found. Hence to reach $M_{rem} \sim 100$ G (measured through $B_z(x, y)$ ($= 4\pi M_{rem}$ as $H = 0$) using MOM (cf. Fig.2)), the magnetic particles in the tape in the vicinity of the laser irradiation spot may have been exposed to $T \geq O(600\ K)$. This observation concurs well with simulations predicting temperatures of $O\ (10^3 K)$ in a localized region[15] around the laser irradiation site. Large fields ($\geq 0.3$ T) stored in the magnetic tape would decay down to few 100 gauss due to localized heating effects. Thus, temperature and relaxation effects obscure the observation of the peak value of the magnetic field recorded on the magnetic tape around the irradiated spot. While local heating of the polymeric tape decays the $M_{rem}$ value, it cannot produce the change in sign of $B_z(x, y)$ as seen in the line scan Fig.2(b).

By determining the curl of $B_z(x, y)$, (where $\vec{B} = B_z(x, y)\hat{k}$), we uncover a symmetry in the complex distribution noted in Fig.2(d). From Fig.4(a), we determine $B_z(x, y = const.)$ and $B_z(x = const., y)$ and take their derivatives in the x and y directions, to obtain the curl of the axial magnetic fields around an irradiated region in Fig.4(b). In Fig.4(b) the light blue region corresponds to maximum values of $\vec{\nabla} \times \vec{B} \approx \hat{i}\, \partial B_z/\partial y \left( \gg \partial B_z/\partial x \right)$ (in x - direction) and orange corresponds to maximum values for $\vec{\nabla} \times \vec{B} \approx \hat{j}\, \partial B_z/\partial x \left( \gg \partial B_z/\partial y \right)$ (in y - direction). The non zero nature of the curl of $\vec{B}_z(x, y)$ implies a circulation ($\vec{\nabla} \times \vec{B} \propto \vec{J}$, where $\vec{J}$ is the current density; displacement current ~ 0 in a plasma). In Fig.4(b) the dotted red circle is a schematic of the vector field represents circulation. We believe our observation of a nonzero circulation associated with $B_z(x, y)$ with its axis perpendicular to the tape surface is due to the circulation of charges in the momentarily generated plasma above the tape. The finite angular momentum generated in the plasma creates a magnetic dipole moment. Information about the perpendicular components of the dipolar magnetic field distributed around the hot expanding plasma gets 'fossilized' or stored in the remnant magnetization of the tape

The circulation of charges in the plasma is akin to vortex motion observed during turbulent flow of rotating fluids. To understand the spatial dependence of the dipolar magnetic field distribution (cf. Fig.2(b)), we approximate the vortex motion in the plasma as an ideal 2D circular sheet of current. A well known feature pertaining to vortices in fluids is that the circulation (vorticity) decays away from the vortex core and the decay are especially strong for low viscosity fluids[16]. A schematic representation of a horizontal cross-section through the plasma is shown in Fig.4(c) inset. Shown as the colored shaded disk is the 2D current sheet circulating in the plasma vortex, whose strengths decay rapidly away from the plasma core (represented as bright near the center and darker around the edges of the disk in the schematic of Fig.4(c)). The current density in the sheet is modeled as $J(r)=J_0 \exp[-(r-\xi_0)/\lambda]$ and $J = 0$ for $r < \xi_0$, where $r$ is the radial coordinate within the plasma, $\xi_0$ is the minimum radius of the current sheet (= radius of the laser spot ~ *10* μm) and λ is radius of the rotating plasma is ~ *100* μm. As shown in the inset of Fig.4(c), λ is the length scale over which the *dipolar* magnetic field distributed (shown as circulating gray arrows) around the hot expanding plasma changes direction (λ ~ *100* μm ,is deduced from the sign change in $B_z(x)$ in Fig.2 (b)). For distance $r > \lambda$, using an Amperian loop shown as the gray circulating arrow in Fig.4(c) inset, the axial component of the *dipolar* magnetic field on z = 0 plane, is $B_z(r \to \infty) \sim \dfrac{\mu_0 \lambda J_0 \exp(\xi_0/\lambda)}{2\pi r}$. From above we approximate that for r >> λ, $B_z(r) \propto r^n$, where *n* is negative. In Figs. 4(c) and 4(d), for the sake of representing the data of Fig. 2(b) on log scale, 100 G has been added to the data in Fig.2(b). Clearly the linear variation in the log-log plot (see blue dotted line in Fig.4(c)), confirms the above approximate $r^n$ variation expected for the *dipolar* field around the plasma. At $r < \lambda$, using Biot-Savart's law we get $B_z(r \to 0) \sim \frac{\mu_0 \lambda J_0}{2\,r} \exp\left[-(r-\xi_0)/\lambda\right]\left[\exp(r/\lambda)-1\right]$. For $r \ll \lambda$, using $\left[\exp(r/\lambda)-1\right] \approx r/\lambda$, we find $|\ln B_z(r)| \propto r$, which is confirmed via a linear variation (dotted red line) observed in the semi-log plot in Fig.4(d). Using our previously estimated lower bound axial field of 0.3 T at $r = 200$ μm and with the $B_z(r \to \infty)$ expression, we evaluate $\mu_0 \lambda J_0 \exp(\xi_0/\lambda) \sim 3 \times 10^{-4}\, T-m$. Using this we get $B_z(r \to 0) \sim \frac{3 \times 10^{-4}}{2\lambda} \sim 1.5\, T$. The estimated value depends on our approximation of the size of the hot expanding plasma, $\lambda \sim 100$ μm, which is an overestimation. We point out that this estimate of the field at the centre of the irradiated spot is very conservative; generated fields could be much larger than this.

In conclusion, we have presented clear evidence for giant axial magnetic fields under resonance absorption of intense, pulsed laser light in solid dense plasmas. We present the first high resolution (micron scale) spatial maps of these giant fields using high sensitivity MOM. Our study reveals broken continuous rotational symmetry in the plasma due to the presence of angular momentum associated with a vortex of circulating charges in the plasma. We have analyzed our results with a model for the plasma vortex. We believe that our work would stimulate future theoretical attempts to understand the instabilities in the plasma which generates the vortex and subsequently large axial magnetic fields. It is possible that the well known massive azimuthal fields in some way couple to the plasma to generate the angular momentum in the plasmas associated with intense laser matter interactions. It may also prove interesting and rewarding to draw analogies of the plasma circulation described here with other well known vortex phenomena.

S. S. Banerjee would like to acknowledge the support from Prof. A. K. Grover, Prof. Eli Zeldov. SSB also acknowledges,the funding received from Director, IIT Kanpur and CSIR, India, No. 03(1066)/06/EMR-II and DST India, No. SR/S2/CMP-17/2005.

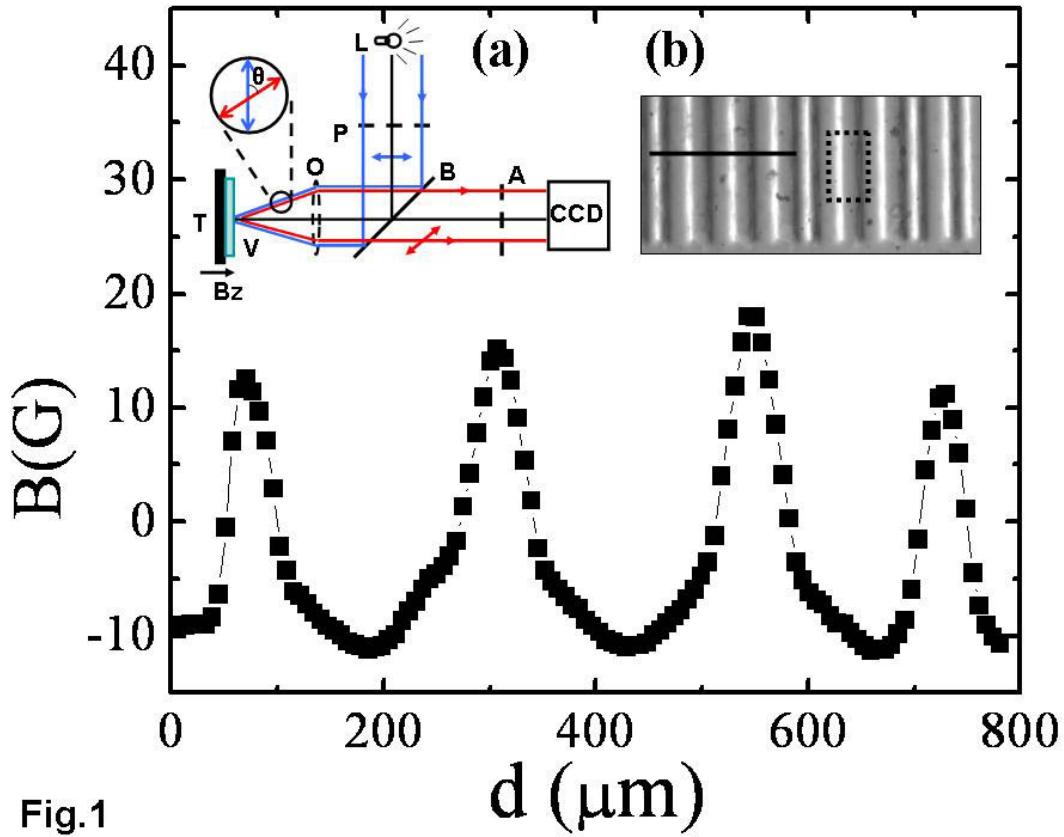

Fig.1. $B_z(x)$ across a region of a prerecorded magnetic tape. (a) Schematic of MOM setup consisting of, lamp (L), linear polarizer (P), beam splitter (B), objective (O), MOM indicator film (V), tape (T), analyzer (A). Also indicated is the local magnetic field direction (Bz). The blow up shows the state of linear polarization in the incident (blue) and reflected beam (red). (b) MOM image of a prerecorded magnetic tape.

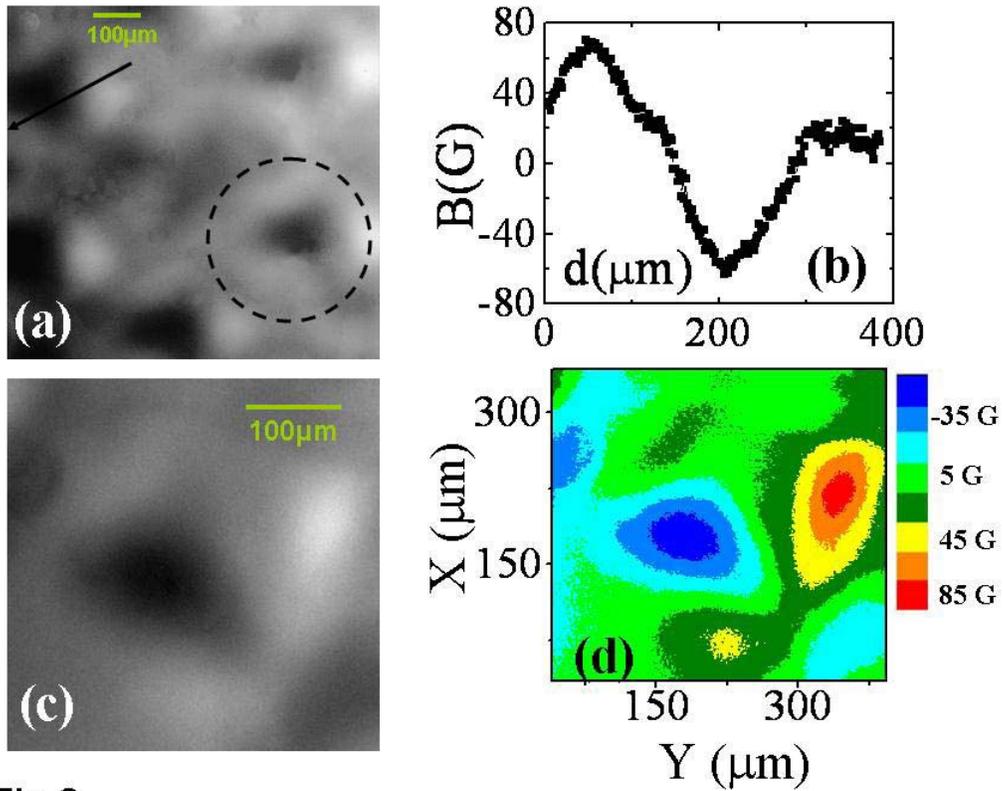

**Fig.2**

Fig.2. (a) MOM image of six locations on the irradiated tape. (b) Behavior of $B_z$ (across the solid line in Fig. 2(a)). (c) Magnified MOM image of a single irradiated spot (refer dotted black circled region in Fig.2(a)). (d) Contour map of $B_z(x, y)$ around the spot shown in (c).

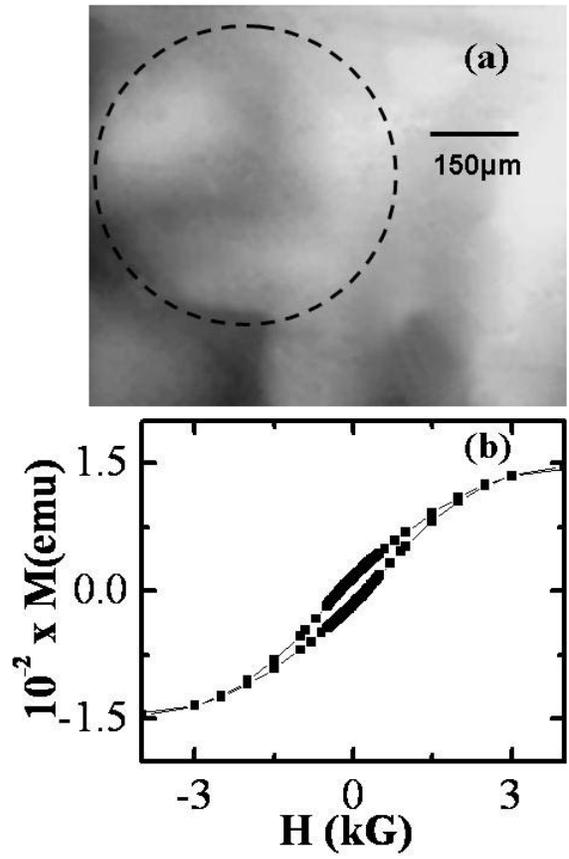

Fig.3 (a) Effect of terawatt laser pulse irradiation on a prerecorded magnetic tape (refer to text for details) (b) M-H hysteresis loop of the tape (cf. text for details)

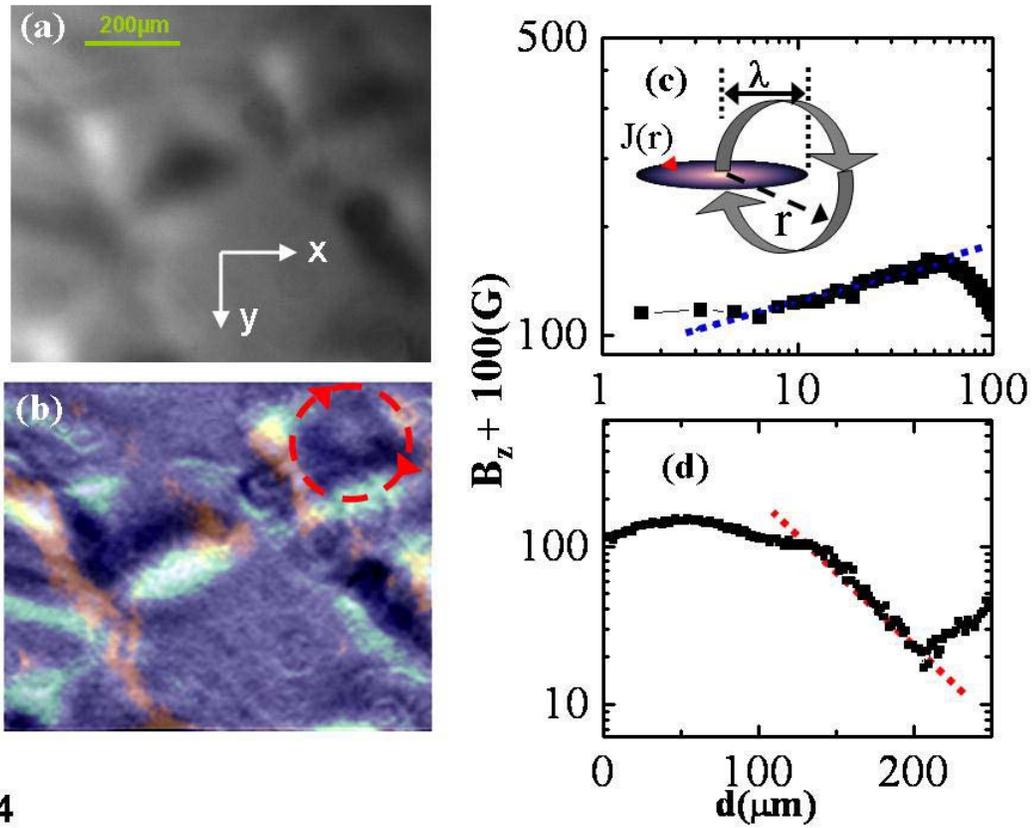

Fig.4

Fig.4 (a) MOM image of irradiated regions on the tape. (b) derivative of the MO image shown in (a) (cf. text). The red dashed circle schematically represents circulation of azimuthal currents responsible for the axial fields. (c) $Log(B_z + 100\ G)$ vs $Log(d)$ variation far away from the laser irradiated spot. Inset shows the schematic of 2D current sheet, J(r) (a model for horizontal cross section of plasma vortex) (d) $Log(B_z + 100\ G)$ vs $d$ variation close to irradiation spot.